\documentclass[11pt]{article}
\usepackage{amsmath}
\usepackage{amssymb}
\usepackage{amscd}
\usepackage{bm}
\usepackage[utf8]{inputenc}
\usepackage{amsfonts}
\usepackage{eufrak}
\usepackage{hyperref}
\usepackage{theorem}
\usepackage{xspace}

\newcommand{\Tr}{{\rm Tr}}

\font\teneusm=eusm10 
\font\seveneusm=eusm7 
\font\fiveeusm=eusm5
\newfam\eusmfam
\textfont\eusmfam=\teneusm \scriptfont\eusmfam=\seveneusm
\scriptscriptfont\eusmfam=\fiveeusm

\font\tencmmib=cmmib10 \skewchar\tencmmib='177
\font\sevencmmib=cmmib7 \skewchar\sevencmmib='177
\font\fivecmmib=cmmib5 \skewchar\fivecmmib='177
\newfam\cmmibfam
\textfont\cmmibfam=\tencmmib \scriptfont\cmmibfam=\sevencmmib
\scriptscriptfont\cmmibfam=\fivecmmib

\font\teneurm=eurm10 
\font\seveneurm=eurm7 
\font\fiveeurm=eurm5
\newfam\eurmfam
\textfont\eurmfam=\teneurm \scriptfont\eurmfam=\seveneurm
\scriptscriptfont\eurmfam=\fiveeurm

\font\teneufm=eufm10 
\font\seveneufm=eufm7 
\font\fiveeufm=eufm5
\newfam\eufmfam
\textfont\eufmfam=\teneufm \scriptfont\eufmfam=\seveneufm
\scriptscriptfont\eufmfam=\fiveeufm

\begin{document}

\begin{titlepage}
\title{{\bf Deformed Kac-Moody Algebra and Noncommutative Fermi Theory in Two-Dimensions} }
\author{M.W.AlMasri\footnote{walidalmasri@usim.edu.my} and M.R.B. Wahiddin\footnote{mridza@usim.edu.my}\\
{\small Cybersecurity and  Systems Unit, Universiti Sains Islam Malaysia},\\ \small  Bandar Baru Nilai, 71800 Nilai, Negeri
	Sembilan, Malaysia\\  
 } 
\maketitle

\abstract{ \noindent Starting from noncommutative Fermi theory in two-dimensions, we construct a  deformed Kac-Moody algebra between its vector and Chiral currents .  The higher-order corrections to the deformed Kac-Moody algebra are explicitly  calculated. We observe  that the ordinary  Schwinger terms are not affected by noncommutativity.  Finally we conclude that the deformed Kac-Moody algebra can be given  in term of ordinary Kac-Moody algebra plus  infinitely many  Lie algebraic structures at each  non-zero power of the antisymmetric  coefficient $\theta$.

}

\end{titlepage}

\section{Introduction}
\label{sec:intro}
Noncommutative field theories has been studied extensively in the previous two decades \cite{Connes,1,2,Seiberg,Schaposnik,Lizzi}. The noncommutativity appears naturally in quantum mechanics when we study the motion of a charged particle moving in two-dimensions under the presence of a constant, perpendicular magnetic field\cite{1,Jackiw} and more broadly in the context of quantum Hall effect\cite{Bellissard,Ezawa,Senthil}. Other studies suggest the appearance  of non-commutativity in the study of open string field theory in the presence of $B$-field\cite{3,Susskind,Sen,4}. Very recently, a possible application in the emerging field of magnetic skyrmions has been reported \cite{Bogdanov,Almasri1,chiral}  \vskip 5mm
It is a well-established fact that the quantum corrections to the classical action 
breakdown the gauge symmetry which is referred to as " anomaly". The non-commutative counterpart of Chiral anomaly and its ramifications has  been studied in many papers ( for example, see \cite{5,Martin1,6,21,Martin,almasri}). \vskip 5mm  Noncommutative field theories in two-dimensions have attracted a lot of attention during the last couple of years; the general structure of gauge  theories can be read from \cite{7}, the formulation of two-dimensional noncommutative gravity was given in \cite{8}. The noncommutative counterpart of other quantum regimes in two-dimensions like Sine-Gordon model and Thirring model were presented in \cite{9}and \cite{10}respectively. \vskip 5mm
The deformation of the conformal symmetry in two-dimensions  and the deformed Kac-Moody algebra  has been studied before \cite{12,13,14}.
In \cite{14}, Balachandran and his collaborators investigated the deformed Kac-Moody algebra from a general perspective, and they ensure that there are two ways for deforming the Kac-Moody algebra: the first deformation is obtained by deforming the oscillators while  the second is obtained directly by deforming the generators of Kac-Moody algebra . These approaches lead to different deformations of the Kac-Moody algebra. It is also shown \cite{12} that by twisting the commutation relations between the creation and annihilation operators, the conformal invariance can still be  maintained in the two-dimensional Moyal plane.\vskip 5mm

In this paper, we continue along these lines and show how to obtain a deformed\footnote{"deformed"throughout this paper means noncommutatively deformed.}
Kac-Moody algebra starting from the massless noncommutative Fermi theory in two-dimensions. We also  calculate the higher-order corrections of the deformed Kac-Moody algebra in powers of the antisymmetric tensor $\theta_{\mu\nu}$.  \vskip 5mm
This article is organised as follows: in section \ref{NCKM}, we give a general derivation of the deformed Kac-Moody algebra starting from a massless noncommutative Fermi theory in two-dimensions. In section \ref{firstorder}, we calculate the higher-order corrections to the deformed Kac-Moody algebra. The 
paper ends with a conclusion and an appendix  about the 
properties of the Moyal $\star$-product. 
\section{The noncommutative deformation of Kac-Moody algebra }\label{NCKM}
Affine Lie algebra is an infinite-dimensional Lie algebra constructed out of a finite-dimensional simple Lie algebra $\mathfrak{g}$ over $\mathbb{C}$ whose generators obey the commutation relations \cite{15}, \cite{16} 
\begin{equation}
[J^{A},J^{B}]= f^{AB}_{C}\;J^{C}
\end{equation}
where $f^{AB}_{C}$ are the antisymmetric structure constants.
The affine Kac-Moody commutation relations are defined by
\begin{equation} \label{KM-commutator}
[J^{A}_{m},J^{B}_{n}]=f^{AB}_{C}\;J^{C}_{m+n}+m\;c\; \delta_{m+n,0}\delta^{AB}
\end{equation}
Where  $\{J^{A}_{m}:m\; \in \; \mathbb{Z}\}$ are the corresponding Laurent modes and $c$ is the central charge. \vskip 5mm
One way to visualize the algebra is to define the generators on $\mathbb{S}^{1}$=$\{z\; \in \; \mathbb{C}: 
|z|=1\}$ parametrised by $\theta\; \in \;[0,2\pi)$, therefore we can  expand the generators as 
\begin{equation}
J^{A}(\theta)=\sum_{m\;\in \; \mathbb{Z}} \;e^{im\theta}\;J^{A}_{m}
\end{equation}
We plug this expansion in the general Kac-Moody commutator (\ref{KM-commutator}) . After some straightforward calculations we find  the following  relations between the currents 
\begin{equation}
[J^{A}(\theta_{1}),J^{B}(\theta_{2})]=2\pi\;f^{AB}_{C}\;J^{C}(\theta_{1})\; \delta(\theta_{1}-\theta_{2})-2\pi i \;c \; \delta^{AB}\; \delta^{'}(\theta_{1}-\theta_{2})
\end{equation}
where we have used the identities $2\pi\;\delta(\theta)=\sum_{m\; \in \; \mathbb{Z}}\;e^{im\theta}$ and $-2\pi i \delta(\theta)=\sum_{m\;\in \; \mathbb{Z}}\;m\; e^{im\theta}$.\\
The term $2\pi i \;c \; \delta^{AB}\; \delta^{'}(\theta_{1}-\theta_{2})$ is  called a " Schwinger term". \\

From physical perspective,  the  Kac-Moody algebra appears for example in the context of  non-Abelian bosonization  in two dimensions \cite{Witten} and in the study of $SU(2)$ chiral spin currents algebra in the Luttinger-Tomonaga liquid \cite{Fradkin}. \vskip 5mm
Our main purpose in this paper is to construct a similar  algebra starting from the noncommutative Fermi theory in two dimensions. \vskip 5mm
We adopt the following conventions throughout this paper :$\gamma^{0}=\sigma^{1}\; , \gamma^{1}=i\sigma^{2}$ and the chirality operator in two-dimensions is $\gamma_{3}=\gamma^{1}\gamma^{0}=i\sigma^{2}\sigma^{1}=\sigma^{3}$ , where $\sigma^{\alpha}$ stands for the  Pauli matrices. \vskip 5mm
The Euclidean version of the theory is obtained  readily after the  Wick rotation ($x^{0}\rightarrow -ix^{2}$), that also  changes the signature of the metric from $(+,-)$ to $(-,-)$ and $\gamma^{o}\rightarrow-i\gamma^{2}$. The Wick rotation of the gauge potential is given by $A_{0}(x)\rightarrow iA_{2}(x)$.\vskip 5mm
The massless Fermi action in  two-dimensional noncommutative Euclidean space-time is given by 
\begin{equation}
S=\int d^{2}x \;  \overline{\psi}(x) \star i\gamma^{\mu}( \partial_{\mu}-i\; \hat{A}_{\mu}^{a}\; T^{a}) \star  \psi(x)                                  .
\end{equation}
$\psi$ is the Fermionic field, $\gamma_{\mu}$'s are the gamma matrices in two-dimensions, $\hat{A}_{\mu}$ is a one-form noncommutative gauge field. The covariant derivative is defined as $D_{\mu}=\partial_{\mu}-i\hat{A^{a}_{\mu}}(x)T^{a}$ and $\star$ is the star-product ( see   appendix  \ref{moyal} for more details on the  $\star$-product) which we define it as \begin{equation}
f(x)\;\star\; g(x)= e^{i\frac{\theta_{\mu\nu}}{2}\; \frac{\partial}{\partial{\zeta_{\mu}}}\;\otimes \frac{\partial}{\partial{\eta_{\nu}}}}\; f(x+\zeta)\; g(x+\eta) \mid_{\zeta=\eta=0}          .
\end{equation}

Then the corresponding path integral is 
\begin{equation}
\mathcal{Z}=\int \mathcal{D}\overline{\psi}\;\mathcal{D}\psi \; e^{S}
\end{equation}
where $S$ is the action functional. \vskip 5mm
The noncommutative Chiral-vector and vector currents are defined in two-dimensions respectively as
\begin{eqnarray}
\hat{j^{a\;\mu}_{3}}(x)=\overline{\psi}(x)\; \gamma^{\mu}\gamma_{3}T^{a}\;\star \psi(x), \\
\hat{j^{a\;\mu}}(x)=\overline{\psi}(x)\; \gamma^{\mu}T^{a}\star \psi(x).
\end{eqnarray}
Under chiral transformation with an infinitesimal $\beta^{a}(x)$ parameter , we can  write the chirally rotated spinors as 
\begin{eqnarray}
\psi^{'}(x)=\mathrm{exp}[i\beta^{a}(x) T^{a} \gamma_{3}]\star\; \psi(x),\\
\overline{\psi^{'}}(x)=\overline{\psi}(x)\star\; \mathrm{exp}[i\beta^{a}(x) T^{a} \gamma_{3}],
\end{eqnarray}
where $\gamma_{3}$ is the chirality operator in two-dimensions.\\ 
After this chiral transformation, the path integral becomes 
\begin{equation}
\mathcal{Z}^{'}=\int \mathcal{D}\overline{\psi}\;\mathcal{D}\psi\; \mathbb{J}\; \star \mathrm{exp}[\int d^{2}x\; \big(\overline{\psi}\star i\gamma^{\mu}(\partial_{\mu}-i\hat{A}^{a}_{\mu} T^{a})\star \psi+ \beta^{a}(x)\star D_{\mu}\hat{j}_{3}^{a\; \mu}(x)\big)]
\end{equation}
$\mathbb{J}$ is the Jacobian of the transformation. We know from \cite{10} that the noncommutative anomaly in two-dimensions is  $\epsilon^{\mu \nu}\hat{F}_{\mu \nu}(x)$ up to a constant , where $\hat{F}_{\mu \nu}$ is the noncommutative two-forms field strength\footnote{ The noncommutative field strength is the  ordinary two-forms field strength plus some terms in order of $\theta$ which we call it the antisymmetric constant tensor\cite{3,17,18}.The first order corrections to the one-form noncommutative vector field and two-forms noncommutative field strength are respectively 
	\begin{equation}
	A^{1}_{\gamma}=-\frac{1}{4}\theta^{\kappa\lambda}\big\{A_{\kappa},\partial_{\lambda}A_{\gamma}+F_{\lambda\gamma}\big\}
	\end{equation}
	
	\begin{equation}
	F^{1}_{\gamma\rho}=-\frac{1}{4}\theta^{\kappa\lambda}\big( \big\{A_{\kappa},\partial_{\lambda}F_{\gamma\rho}+D_{\lambda}F_{\gamma\rho}\big\}-2\big\{F_{\gamma\kappa},F_{\rho\lambda}\big\} \big)
	\end{equation}}; thus the divergence of Chiral current is 
\begin{equation} \label{anomaly}
\langle D_{\mu}\; \hat{j^{a\; \mu}_{3}}(x) \rangle = \partial_{\mu}\langle \hat{j^{a\;\mu}_{3}}(x)\rangle + f^{abc}\;\hat{A^{b}_{\mu}}(x)\star \langle\hat{j^{c\mu}_{3}}(x)\rangle =- \frac{i}{4\pi}\epsilon^{\mu\nu}\;\hat{F^{a}_{\mu\nu}}(x)
\end{equation}
Here we assumed that the gauge group generators satisfy  the normalization condition  $\Tr(T^{a}T^{b})=\frac{1}{2}\delta^{ab}$ and the averaged quantities are defined by
\begin{equation}
\langle \mathcal{O}(x) \rangle=\int \mathcal{D}\overline{\psi}\mathcal{D}\psi \; \mathcal{O}(x)\star \mathrm{exp}[\int d^{2}x\; \overline{\psi}\star i\gamma^{\mu}(\partial_{\mu}-i\hat{A^{a}_{\mu}}T^{a})\star \psi]  .
\end{equation}
Under vector-like transformations
\begin{eqnarray}
\psi^{'}=\mathrm{exp}[i\; \alpha^{a}(x)\;T^{a}]\star \psi(x),\\
\overline{\psi^{'}}=\overline{\psi}(x)\star \mathrm{exp}[-i\; \alpha^{a}(x)\;T^{a}]
\end{eqnarray}
we obtain a similar relation for the vector current : 
\begin{equation}
\langle D_{\mu}\; \hat{j^{a\; \mu}}(x) \rangle = \partial_{\mu}\langle \hat{j^{a\;\mu}}(x)\rangle + f^{abc}\;\hat{A^{b}_{\mu}}(x)\star \langle\hat{j^{c\mu}}(x)\rangle =0  .
\end{equation}
We normalize the antisymmetric tensor according to $\epsilon^{12}=1$,  note that the Dirac-delta function $\delta(x-y)$  and the  $\epsilon^{\mu\nu}$ symbols are defined as the ordinary commutative case. 

\vskip 5mm 

In the massless case, as a special property of living in two-dimensions, we may write the following  relation between the Chiral and vector currents: 
\begin{equation}
\hat{j^{a\mu}_{3}}(x)=\overline{\psi}(x)\;\star\gamma^{\mu}T^{a}\gamma_{3}\psi(x)=-\epsilon^{\mu\nu}\;\overline{\psi}(x)\;\star T^{a}\gamma_{\nu}\psi(x)=-\epsilon^{\mu\nu}\;\hat{j^{a}_{\nu}}(x)  .
\end{equation}

We functionally differentiate (\ref{anomaly}) with respect to $\hat{A^{b}_{\nu}}(y)$ and send $\hat{A^{b}_{\nu}}$ to zero:
\begin{eqnarray}\label{diff1}
\frac{\delta}{\delta\hat{A^{b}_{\nu}}(y)} \langle D_{\mu}\hat{j^{a\;\mu}_{3}}(x)\rangle=\partial_{\mu}\langle \mathcal{T}\hat{j^{a\;\mu}_{3}}(x)\star \hat{j^{b\;\nu}}(y) \rangle + f^{abc}\; \delta^{2}(x-y)\star \langle \hat{j^{c\;\nu}_{3}}(x) \rangle \\ \nonumber
=-\frac{i \epsilon^{\mu\nu}\;\delta_{ab}}{2\pi}\partial{\mu}\delta^{2}(x-y) .
\end{eqnarray}
In the same manner, we obtain the following relation for the noncommutative  vector current :
\begin{eqnarray}\label{diff2}
\frac{\delta}{\delta\hat{A^{b}_{\nu}}(y)} \langle D_{\mu}\hat{j^{a\;\mu}}(x)\rangle=\partial_{\mu}\langle \mathcal{T}\hat{j^{a\;\mu}}(x)\star \hat{j^{b\;\nu}}(y) \rangle + f^{abc}\; \delta^{2}(x-y)\star \langle \hat{j^{c\;\nu}}(x) \rangle=0
\end{eqnarray}
We apply the Bjorken-Johonson-Low (BJL) prescription to convert the time-ordering $\mathcal{T}$ in the Lorentz covariant calculations  to the normal time-ordering $T$ which is not  necessarily Lorentz covariant.\footnote{Suppose $\hat{X},\hat{Y}$ some noncommutative quantities then the BJL prescription can be build as follows 
	\begin{equation} \nonumber
	\lim_{k_{2} {\to} \infty}\int d^{2}x\; e^{ikx}\langle \mathcal{T}\hat{X}(x)\star\hat{Y}(0)\rangle=0
	\end{equation}
	\begin{eqnarray}\nonumber
	\lim_{k_{2} {\to} \infty}\int d^{2}x\; e^{ikx}\langle T\hat{X}(x)\star \hat{Y}(0)\rangle=\int d^{2}x e^{ikx}\langle
	\mathcal{T}\hat{X}(x)\star \hat{Y}(o)\rangle-\lim_{k_{2} {\to} \infty}\int d^{2}x\; e^{ikx}\langle \mathcal{T}\hat{X}(x)\star\hat{Y}(0)\rangle
	\end{eqnarray}} 
For more details about the usage of BJL prescription in the noncommutative case,see \cite{22} where this prescription is applied in both higher-derivatives scalar field and noncommutative scalar field theory.

The equations (\ref{diff1}) and  (\ref{diff2}) can be written as 
\begin{equation}\label{1}
\partial_{\mu}\langle T \hat{j^{a\;\mu}_{3}}(x)\star \hat{j^{b2}}(y)\rangle+f^{abc}\langle \hat{j^{c\;2}}\rangle \star \delta^{2}(x-y)=-\frac{i\delta_{ab}}{2\pi}\partial_{1}\delta^{2}(x-y)
\end{equation}

\begin{equation}\label{2}
\partial_{\mu}\langle T \hat{j^{a\;\mu}_{3}}(x)\star \hat{j^{b1}}(y)\rangle+f^{abc}\langle \hat{j^{c\;1}}\rangle \star \delta^{2}(x-y)=0
\end{equation}

\begin{equation}\label{3}
\partial_{\mu}\langle T \hat{j^{a\;\mu}}(x)\star \hat{j^{2}}(y)\rangle+f^{abc}\langle \hat{j^{c\;2}}\rangle \star \delta^{2}(x-y)=0
\end{equation}
Since the left hand side of the equation (\ref{2}) is defined in terms of a time-ordered product so we eliminate the right-hand side and set it to zero.
We use the relations $\hat{j^{b\;1}}(y)=i\hat{j^{b\;2}_{3}}(y)$,$\hat{j^{c\;1}_{3}}(x)=i\hat{j^{c\;2}}(x)$ and $\partial_{\mu}\hat{j^{a\;\mu}}=\partial_{\mu}\hat{j^{a\;\mu}_{3}}=0$ to write the equations (\ref{1}), (\ref{2}) and \ref{3} as
\begin{equation}
[\hat{j^{a\;2}_{3}}(x),\hat{j^{b\;2}}(y)]_{\star}\;\star \delta(x^{2}-y^{2})+f^{abc}\hat{j^{c\;2}}(x)\star \delta^{2}(x-y)=-\frac{i\delta_{ab}}{2\pi}\partial_{1}\delta^{2}(x-y)
\end{equation}

\begin{equation}
[\hat{j^{a\;2}_{3}}(x),\hat{j^{b\;2}_{3}}(y)]_{\star}\;\star \delta(x^{2}-y^{2})+f^{abc}\hat{j^{c\;2}}(x)\star \delta^{2}(x-y)=0
\end{equation}

\begin{equation}
[\hat{j^{a\;2}}(x),\hat{j^{b\;2}}(y)]_{\star}\;\star \delta(x^{2}-y^{2})+f^{abc}\hat{j^{c\;2}}(x)\star \delta^{2}(x-y)=0
\end{equation}
Note that $\partial_{\mu}\hat{j^{a\;\mu}_{3}}\neq 0$ in the massive Fermi theory. \vskip 5mm
Let us define the following quantities 
\begin{eqnarray}
\hat{j^{a\;2}_{L}}(x)=\frac{1}{2}[\hat{j^{a\;2}}(x)-\hat{j^{a\;2}_{3}}(x)],\;\;\; \;\;
\hat{j^{a\;2}_{R}}(x)=\frac{1}{2}[\hat{j^{a\;2}}(x)+\hat{j^{a\;2}_{3}}(x)],\\
\hat{j^{b\;2}_{L}}(x)=\frac{1}{2}[\hat{j^{b\;2}}(x)-\hat{j^{b\;2}_{3}}(x)],\;\;\; \;\; 
\hat{j^{b\;2}_{R}}(x)=\frac{1}{2}[\hat{j^{b\;2}}(x)+\hat{j^{b\;2}_{3}}(x)],\\ 
\hat{j^{c\;2}_{L}}(x)=\frac{1}{2}[\hat{j^{c\;2}}(x)-\hat{j^{c\;2}_{3}}(x)],\;\;\; \;\; 
\hat{j^{c\;2}_{R}}(x)=\frac{1}{2}[\hat{j^{c\;2}}(x)+\hat{j^{c\;2}_{3}}(x)] .
\end{eqnarray}

and use them finally  obtain the equal-time commutation relations 
\begin{equation}\label{DKM1}
[\hat{j^{a\;2}_{L}}(x),\hat{j^{b\;2}_{L}}(y)]_{\star}=-f^{abc}\hat{j^{c\;2}_{L}}(x)\star \delta(x^{1}-y^{1})+\frac{i\delta_{ab}}{4\pi}\partial_{1}\delta(x^{1}-y^{1}) ,
\end{equation}

\begin{equation}\label{DKM2}
[\hat{j^{a\;2}_{R}}(x),\hat{j^{b\;2}_{R}}(y)]_{\star}=-f^{abc}\hat{j^{c\;2}_{R}}(x)\star \delta(x^{1}-y^{1})-\frac{i\delta_{ab}}{4\pi}\partial_{1}\delta(x^{1}-y^{1}).
\end{equation}

Comparing \ref{DKM1}and \ref{DKM2} with the deformed Kac-Moody algebra obtained from deforming oscillators \cite{14}, we found in contrast the central charge ( Schwinger) term to be unaffected by non-commutativity. This can be explained using the properties of $\star$-product, namely the fact that Dirac delta function in the noncommutative case is defined as $\int f(x)\star \delta^{2}(x-y)\; d^{2}x= f(y)$ for test function $f(x)$ which is identical to the commutative case plus the fact that $\partial_{\mu}\star\delta^{2}(x-y)= \partial_{\mu}\delta^{2}(x-y)$. Thus one can view the deformed Kac-Moody algebra introduced  in \cite{14} as the most general deformation for all $SU(N)$ Kac-Moody algebras while in our work  is for specific $SU(2)$  Kac-Moody algebra with chiral and vector currents from  noncommutative  Fermi theory in two-dimensions.

\section{Deformed Kac-Moody algebra to all orders in $\theta$}\label{firstorder}
In order to make a strong conclusion about the nature of deformed Kac-Moody algebra \ref{DKM1} and \ref{DKM2}, it is interesting to explore the higher-order corrections  in  the antisymmetric noncommutative tensor $\theta^{\mu\nu}$. 

The expansion of \ref{DKM1}  gives  
\begin{align}
[\hat{j^{a\;2}_{L}}(x),\hat{j^{b\;2}_{L}}(y)]_{\star}= [\hat{j^{a\;2}_{L}}(x), \hat{j^{b\;2}_{L}}(y)]- \sum_{n=1}^{\infty}\frac{f^{abc}}{n!}\hat{j^{c\;2}_{L}}(x) (\frac{1}{2}\theta^{\mu\nu} \partial^{\leftarrow}_{\mu} \partial^{\rightarrow}_{\nu})^{n}  \delta(x^{1}-y^{1})
\end{align}
where $[\hat{j^{a\;2}_{L}}(x), \hat{j^{b\;2}_{L}}(y)]$ has the ordinary Kac-Moody algebra structure and equals to 
\begin{equation}
[\hat{j^{a\;2}_{L}}(x), \hat{j^{b\;2}_{L}}(y)]= -f^{abc} \hat{j^{c\;2}_{L}}(x) \delta(x^{1}-y^{1})+ \frac{i \delta_{ab}}{4\pi}\partial_{1}\delta(x^{1}-y^{1})
\end{equation}

Analogously we may write the higher-order expansion for \ref{DKM2} as 
\begin{align}
[\hat{j^{a\;2}_{R}}(x),\hat{j^{b\;2}_{R}}(y)]_{\star}= [\hat{j^{a\;2}_{R}}(x), \hat{j^{b\;2}_{R}}(y)]- \sum_{n=1}^{\infty}\frac{f^{abc}}{n!}\hat{j^{c\;2}_{R}}(x) (\frac{1}{2}\theta^{\mu\nu} \partial^{\leftarrow}_{\mu} \partial^{\rightarrow}_{\nu})^{n}  \delta(x^{1}-y^{1})
\end{align}
and the corresponding ordinary Kac-Moody algebra as 
\begin{equation}
[\hat{j^{a\;2}_{R}}(x), \hat{j^{b\;2}_{R}}(y)]= -f^{abc} \hat{j^{c\;2}_{R}}(x) \delta(x^{1}-y^{1})+ \frac{i \delta_{ab}}{4\pi}\partial_{1}\delta(x^{1}-y^{1})
\end{equation}
From the higher-order expansion we note that the deformed Kac-Moody algebra can be written as ordinary Kac-Moody algebra plus infinitely many Lie algebra structures of the form 
\begin{align}
\sum_{n=1}^{\infty}\frac{1}{n!} \hat{j^{a\;2}_{j}}(x)  (\frac{1}{2}\theta^{\mu\nu} \partial^{\leftarrow}_{\mu} \partial^{\rightarrow}_{\nu})^{n}  \hat{j^{b\;2}_{j}}(y)- \sum_{n=1}^{\infty}\frac{1}{n!} \hat{j^{b\;2}_{j}}(y)   (\frac{1}{2}\theta^{\mu\nu} \partial^{\leftarrow}_{\mu} \partial^{\rightarrow}_{\nu})^{n}  \hat{j^{a\;2}_{j}}(x)\\ \nonumber =- \sum_{n=1}^{\infty}\frac{f^{abc}}{n!}\hat{j^{c\;2}_{j}}(x) (\frac{1}{2}\theta^{\mu\nu} \partial^{\leftarrow}_{\mu} \partial^{\rightarrow}_{\nu})^{n}  \delta(x^{1}-y^{1})
\end{align}
where $j=L,R$.

\section{Conclusion}
We have explicitly obtained the deformed Kac-Moody algebra  starting from the two-dimensional noncommutative Fermi theory. These deformations are different from  those given in  paper \cite{14}  where the central charge term is modified  by the noncommutativity for deformations that are  obtained directly from deforming the oscillators. 
In our case the Schwinger term ( i.e. central charge term ) is not affected by the  non-commutativity. 
Finally, we conclude that the  deformed  Kac-Moody algebra in the two-dimensional noncommutative  Fermi theory can be written as ordinary Kac-Moody algebra plus infinitely many embedded Lie algebraic structures. Possible applications of our results can be found in the study of  $SU(2)$ chiral currents in the noncommutative Luttinger-Tomonaga liquid and  non-Abelian Bosonization of massless non-commutative Fermi theory in two dimensions. 
\section*{Acknowledgment} 
One of the authors (M.W.A) is grateful to G. Thompson and K. S. Narain for discussions on quantum anomalies. We thank our referee for the insightful comments and suggestions that improved this paper.   We are grateful to USIM for support. 
\appendix
\section{Moyal $\star$-Product}\label{moyal}
Let $C^{\infty}(\mathbb{R}^{2n})$ be the space of real  smooth functions, 
$f:\mathbb{R}^{2n}\rightarrow \mathbb{C}^{n}$.Then given  $f,g$  $\epsilon$ $C^{\infty}(\mathbb{R}^{2n})$,
the Moyal $\star$-product is defined as 
\begin{equation}
f(x)\;\star\; g(x)= e^{i\frac{\theta_{\mu\nu}}{2}\; \frac{\partial}{\partial{\zeta_{\mu}}}\;\otimes \frac{\partial}{\partial{\eta_{\nu}}}}\; f(x+\zeta)\; g(x+\eta) \mid_{\zeta=\eta=0}
\end{equation}
where $\theta_{\mu\nu}=\theta\epsilon_{\mu\nu}$ is a real anti-symmetric constant, and leads to $C^{\star}$-algebra.  
By expanding the previous formula up to first order in $\theta$, we find that 
\begin{equation}
f\;\star g= f\;q+\frac{i}{2}\; \theta^{\mu\nu}\; \partial_{\mu}f\; \partial_{\nu}g+ \mathcal{O}(\theta^{2}).
\end{equation}
The star product has many properties like 
\begin{equation}
f\; \star g \mid _{\theta}=g\; \star f \mid_{-\theta} .
\end{equation}
Suppose $h$ is another smooth function. Then we have the following cyclic property :
\begin{equation} \label{cyclic}
\int_{-\infty}^{\infty}\; d^{D}x\; ( f \star g \star h)=\int_{-\infty}^{\infty}\; d^{D}x\; ( h \star f \star g)=\int_{-\infty}^{\infty}\; d^{D}x\; ( g \star h \star f) .
\end{equation}
The Leibniz rule holds in the $\star$-product case:
\begin{equation}
\partial_{\mu}(f\star g)=(\partial_{\mu}f)\star g + f\star  (\partial_{\mu}g) .
\end{equation}
The ordinary commutators are generalized in the noncommutative case to the Moyal $\star$-brackets
\begin{equation}
[f,g]_{\star}=f\star\;g - g\star\;f       .
\end{equation}
Then the deformed Jacobi identity would be written as 
\begin{equation}
[f,[g,h]_{\star}]_{\star}+[h,[f,g]_{\star}]_{\star}+[g,[h,f]_{\star}]_{\star}=0 . 
\end{equation}

It is important to notice the fact that any arbitrary  commutator with Moyal $\star$-product involves both ordinary commutators and anti-commutators \cite{20}: 
\begin{equation}
f\star\;g - g\star\;f = [f,g]_{(\star,even)}+ \{f,g\}_{(\star,odd)}
\end{equation}
\begin{equation}
[f,g]_{(\star,even)}=[f,g]+(\frac{i}{2})^{2}\; \theta^{\mu\nu}\theta^{\rho\sigma}[\partial_{\mu}\partial_{\rho}f,\partial_{\nu}\partial_{\sigma}g]+\mathcal{O}(\theta^{4})
\end{equation}
\begin{equation}
\{f,g\}_{(\star,odd)}=\frac{i}{2}\theta^{\mu\nu}\{\partial_{\mu}f,\partial_{\nu}g\}+(\frac{i}{2})^{3}\theta^{\mu\nu}\theta^{\rho\sigma}\theta^{\kappa\lambda} \{\partial_{\mu}\partial_{\rho}\partial_{\kappa}f,\partial_{\nu}\partial_{\sigma}\partial_{\lambda}g\}+\mathcal{O}(\theta^{5})
\end{equation}
We define the  symmetric and anti-symmetric parts of a  $\star$-product as 

\begin{eqnarray}\label{cosformula}
(f\star g)_{s}=\frac{1}{2}(f\star g+ g \star f)=f\;\mathrm{cos}(\frac{1}{2}\overleftarrow{\partial_{\mu}}\theta^{\mu\nu}\overrightarrow{\partial_{\nu}})\;g \\ \simeq \nonumber fg+(\frac{i}{2})^{2}\; \theta^{\mu\nu}\theta^{\rho \sigma}\; \partial_{\mu}\partial_{\rho}f\; \partial_{\nu}\partial_{\sigma} g+\mathcal{O}(\theta^{4}),
\end{eqnarray}

\begin{eqnarray}\label{sinformula}
(f \star g)_{a}=\frac{1}{2}(f \star g-g \star f)=if\;\mathrm{sin}(\frac{1}{2}\overleftarrow{\partial_{\mu}}\theta^{\mu\nu}\overrightarrow{\partial_{\nu}})\;g \\ \nonumber \simeq  (\frac{i}{2}) \theta^{\mu\nu} \; \partial_{\mu} f \partial_{\nu} g+ (\frac{i}{2})^{3}\theta^{\mu\nu}\theta^{\rho\sigma}\theta^{\kappa \lambda} \; \partial_{\mu}\partial_{\rho}\partial_{\kappa}f \; \partial_{\nu}\partial_{\sigma}\partial_{\lambda}g
+\mathcal{O}(\theta^{5}) . 
\end{eqnarray}

\end{document}